\documentclass{appolb}
\usepackage{epsfig}

\begin{document}
\pagestyle{plain}
\newcount\eLiNe\eLiNe=\inputlineno\advance\eLiNe by -1
\title{FEASIBILITY STUDIES FOR \\
THE $\overline{\textrm{P}}$ANDA EXPERIMENT AT FAIR 
\thanks{Presented at the XXXI Mazurian Lakes Conference on Physics,
  Piaski, 30.08-6.09, 2009}
\hspace{1mm}
\thanks{This work is supported by Veni-grant 680-47-120 from the Netherlands 
  Organisation for Scientific Research (NWO), the University of Groningen and the {\it{Helmholtzzentrum f\"ur Schwerionenforschung GmbH}} (GSI), Darmstadt.
}%
}
\author{A.~Biegun
  \address{KVI/University of Groningen, 9747 AA Groningen,  The Netherlands \\
    for the $\overline{\textrm{P}}$ANDA Collaboration}}

\newcommand\fns[1]{\footnotesize{#1}}
\maketitle

\vspace*{-0.9cm}
\begin{abstract}
\hspace*{-0.4cm} $\overline{\verb+P+}$\verb+ANDA+,  
the detector to study Anti$\underline{\textrm{P}}$roton 
$\underline{\textrm{AN}}$nihilations 
at $\underline{\textrm{DA}}$rmstadt, will be installed at the future 
international $\underline{\textrm{F}}$acility for 
$\underline{\textrm{A}}$ntiproton and $\underline{\textrm{I}}$on 
$\underline{\textrm{R}}$esearch (\verb+FAIR+) in Darmstadt, Germany. 
The $\overline{\verb+P+}$\verb+ANDA+  
physics program is 
oriented towards the studies of the strong interaction and hadron structure
performed with the highest quality beam of anti-protons~[1].\newline
\hspace*{0.2cm} 
In the preparation for $\overline{\verb+P+}$\verb+ANDA+ expe\-riments, 
large-scale si\-mu\-lation stu\-dies are being performed 
to validate the performance of all individual detector components and 
to advice on detector optimisation.
The feasibility of the analysis strategies together with the calibration
methods are being studied.\newline
\hspace*{0.2cm} Simulations were carried out using the framework called 
\verb+PandaROOT+~[2], based on \verb+ROOT+ and the 
\verb+Virtual Monte Carlo+ concept~[3]. 
\end{abstract}


\section{Introduction}
\hspace*{-0.4cm} With the $\overline{\verb+P+}$\verb+ANDA+  
physics program 
various questions related to the pro\-perties of the strong interaction 
will be answered. 
This experiment operates in the transition region between 
the perturbative and the non-perturbative QCD regimes and, thereby, will gain 
insight into the mechanism of hadron mass generation and quark confinement.

\hspace*{-0.3cm}The highest quality beam of anti-protons, in terms of intensity and re\-so\-lution,
is the key ingredient for the $\overline{\verb+P+}$\verb+ANDA+  
project.
The momentum range of anti-protons of 1.5$-$15~GeV/c gives access 
to a center-of-mass energy range from 2.2$-$5.5~GeV/c$^{2}$ in
$\overline{\textrm p}\textrm p$ annihilations. The $\overline{\verb+P+}$\verb+ANDA+  
detector 
will measure precisely momenta and scattering angles of both charged 
and neutral particles of the reaction products to detect the complete spectrum of final
states relevant for the $\overline{\verb+P+}$\verb+ANDA+  
physics cases.

\section{The PandaROOT framework}
\hspace*{-0.4cm}The offline software for $\overline{\verb+P+}$\verb+ANDA+,  
\verb+PandaROOT+, 
is an extension of the \verb+FairROOT+ [4] framework 
being designed for the \verb+FAIR+ project.
 The \verb+PandaROOT+ framework, 
being developed by the $\overline{\verb+P+}$\verb+ANDA+
collaboration, is being used for both si\-mulation and data analysis 
of the $\overline{\verb+P+}$\verb+ANDA+
physics stu\-dies. 
It contains tools adapted from high energy physics experiments, 
such as event generators, to produce the reaction of interest 
and its decay chain, 
to ge\-ne\-rate the background from $\overline{\textrm p}\textrm p$ 
 annihilation 
or to study the $\overline{\textrm  p}$A collision.
Also, va\-rious tools allowing to test the reconstruction algorithm, 
and the complete digitization and reconstruction chain, and allowing 
to perform a detailed analysis together 
with the full geometry model are included in this framework.
\verb+PandaROOT+ is highly versatile and enables a user-friendly
interface to various transport engines, like GEANT3,
GEANT4 and Fluka~[5]. 

\section{The $\overline{\textrm{P}}$ANDA setup}
\vspace*{-0.5cm}
\begin{figure}[htb]
\centering
\includegraphics[width=8.cm]{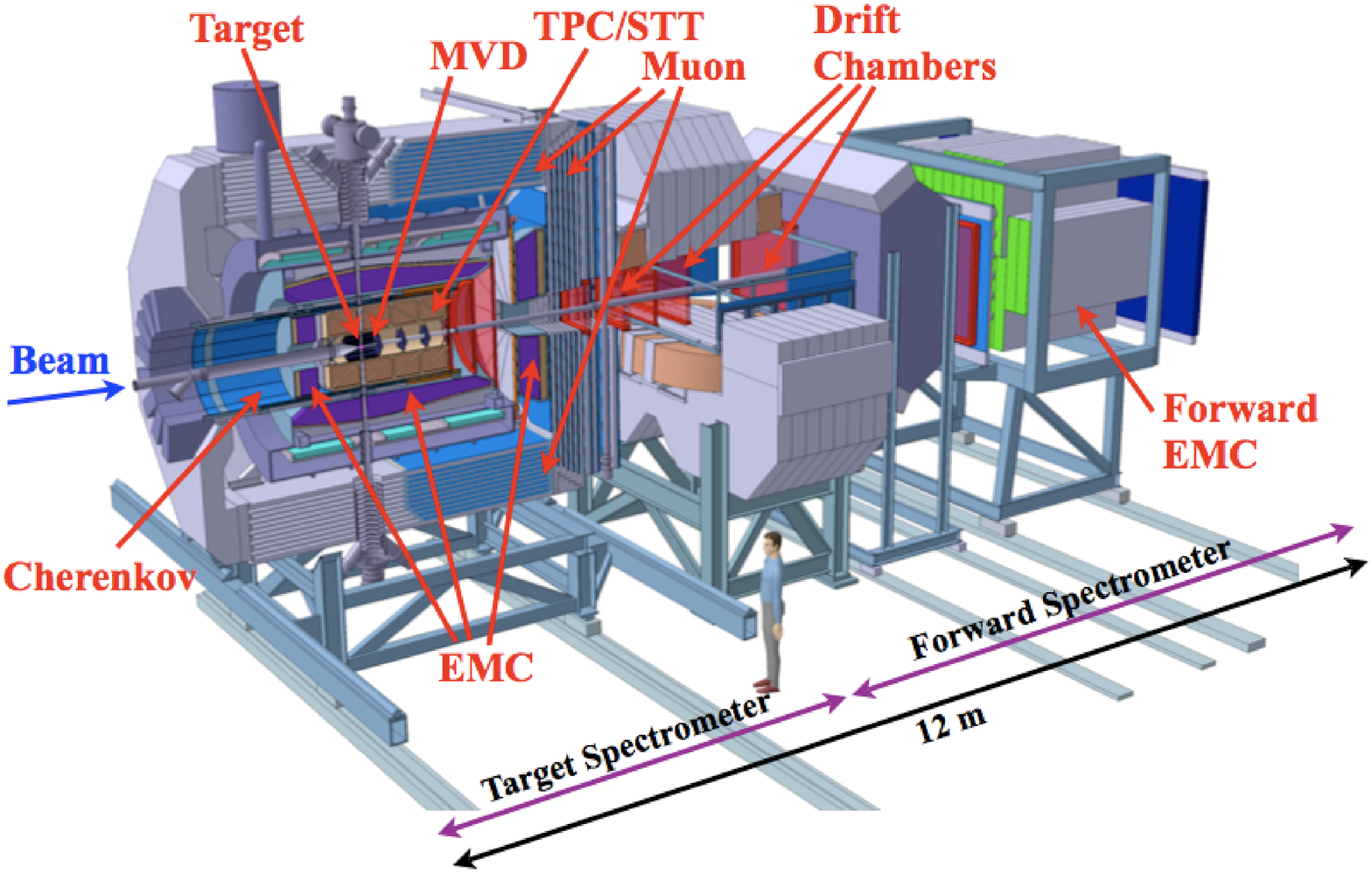}
\caption{\label{fig1}The $\overline{\textrm{P}}$ANDA detection system.}
\end{figure}
\begin{figure}[hb]
\begin{center}
\begin{minipage}[t]{0.495\textwidth}
\includegraphics[width=6.2cm]{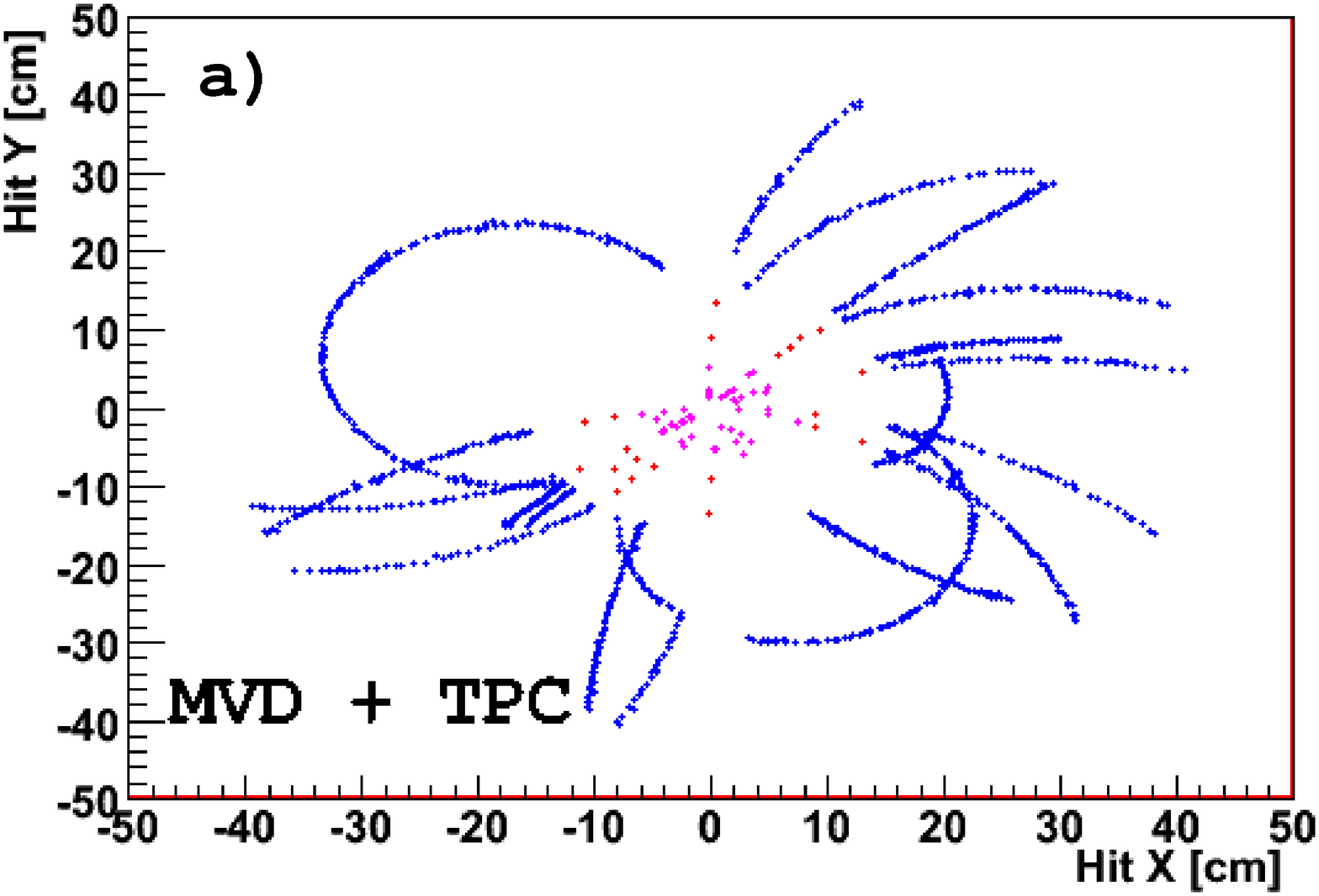}
\end{minipage} \hfill
\begin{minipage}[t]{0.495\textwidth}
\includegraphics[width=6.0cm]{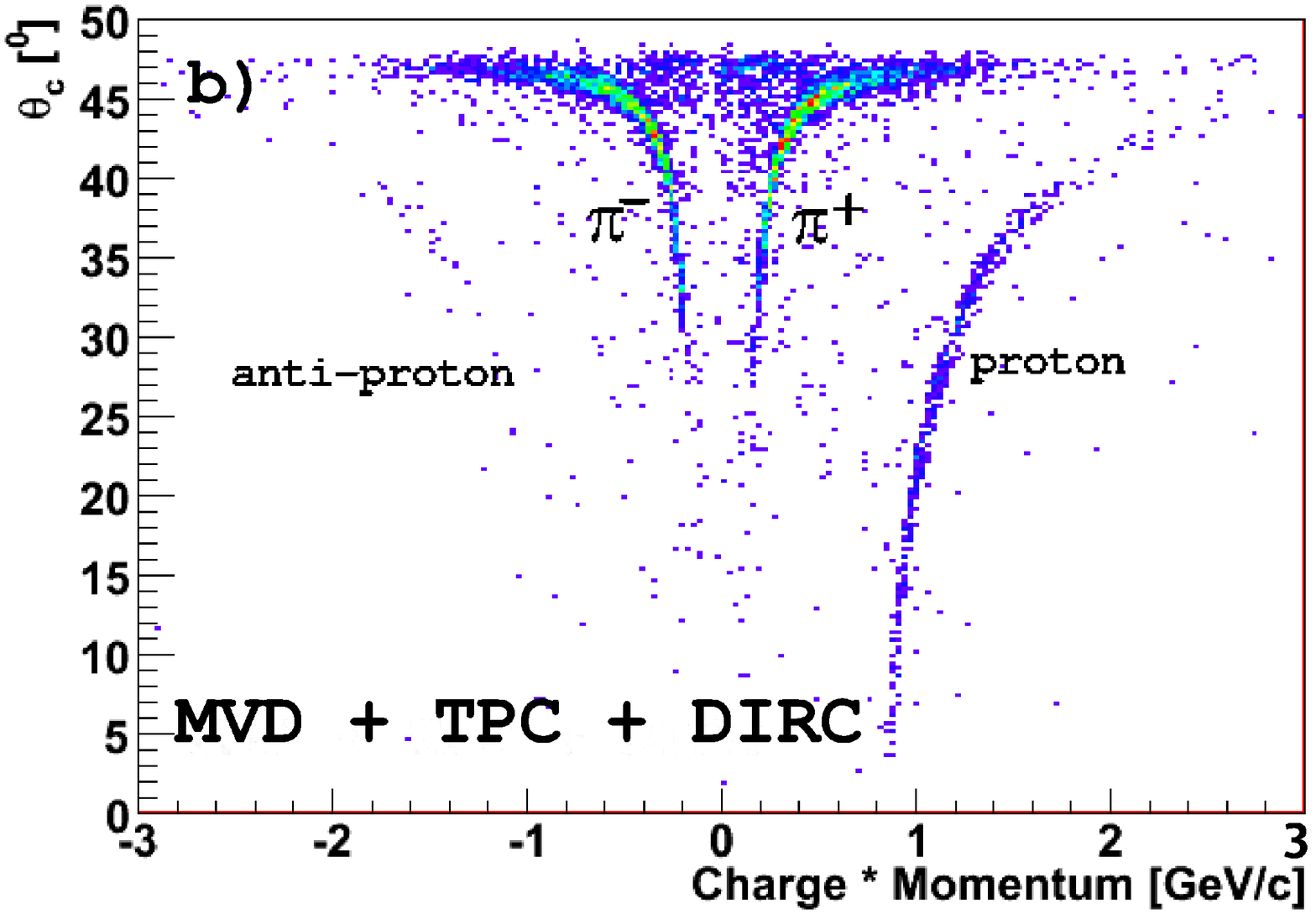}
\end{minipage}
\vspace*{-0.5cm}
\caption{\label{fig2}{\bf{a)}} Recontructed tracks of charged particles from track segments 
in MVD (dots in the middle) and TPC (outer curves) detectors. 
{\bf{b)}} Identification of charged particles with different momenta and charge
using the MVD, TPC and DIRC detectors. 
The $Y$ axis represents the Cherenkov angle $\Theta$. Plots are taken from
S.~Spataro, INFN Torino.}
\end{center}
\end{figure}
\begin{figure}[ht]
\begin{center}
\begin{minipage}[t]{0.495\textwidth}
\vspace*{-120pt}
\includegraphics[width=6.0cm]{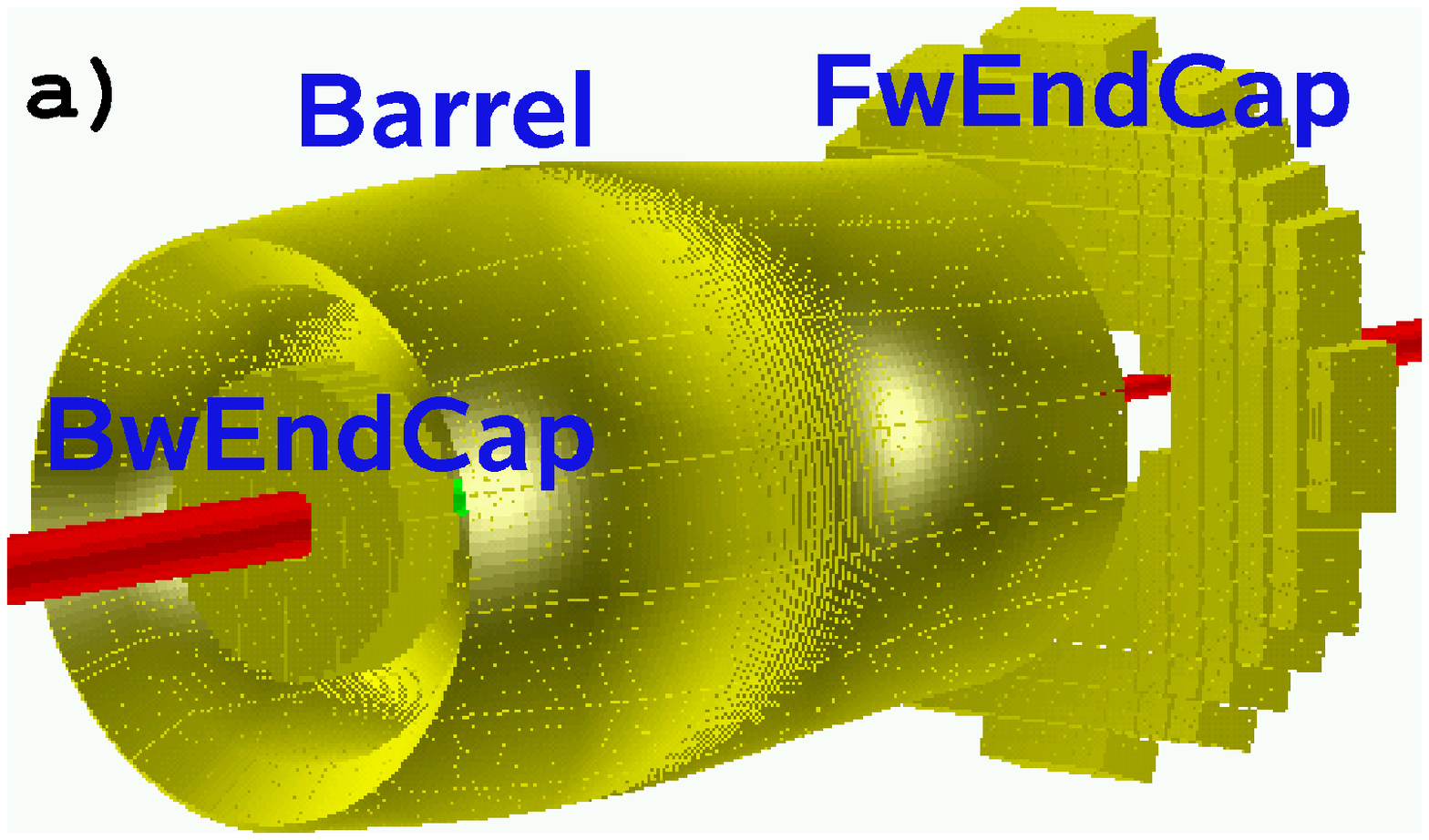}
\end{minipage} \hfill
\begin{minipage}[t]{0.48\textwidth}
\includegraphics[width=6.1cm]{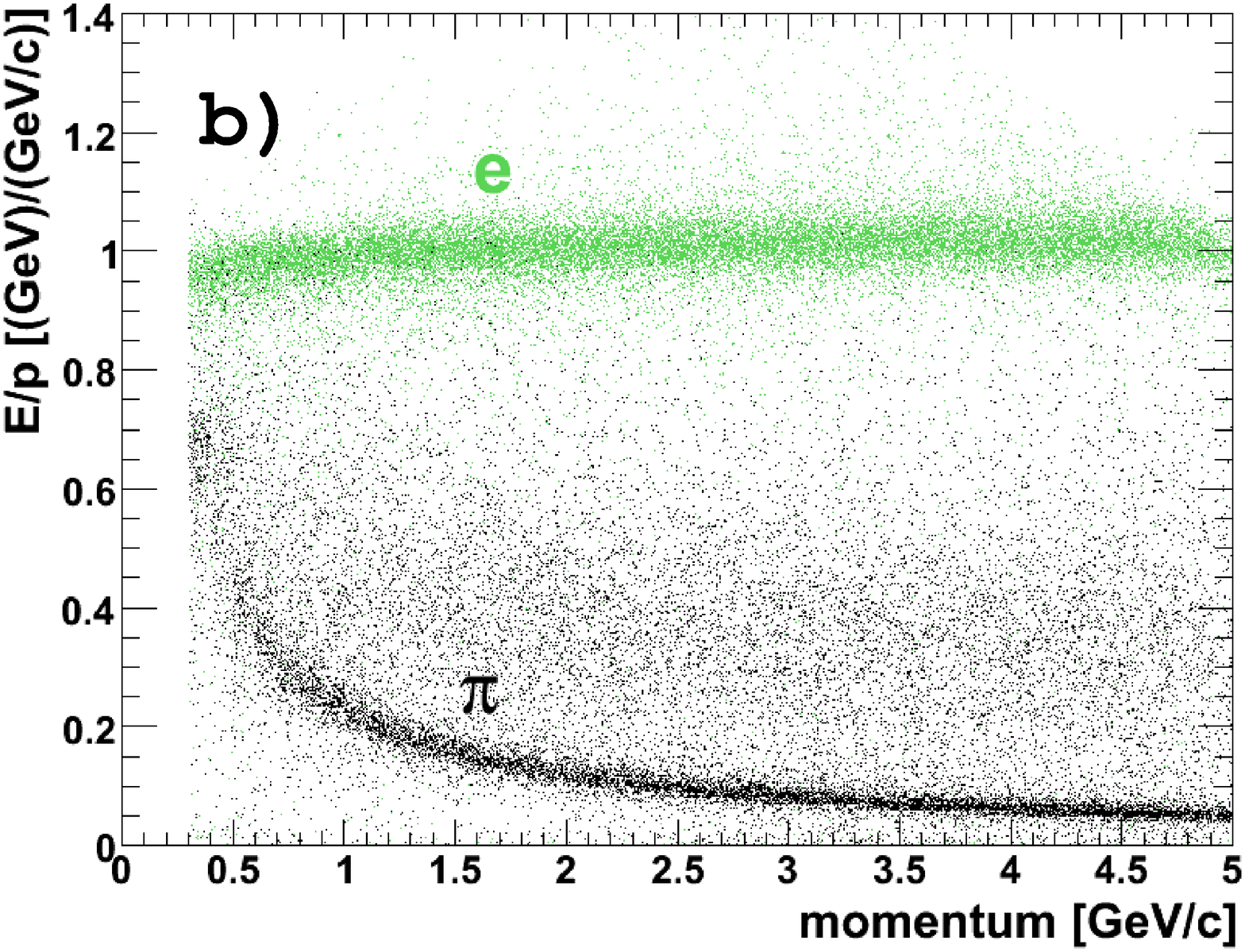}
\end{minipage}
\vspace*{-0.2cm}
\caption{\label{fig3}{\bf{a)}} A 3D view of the electromagnetic calorimeter in the target spectrometer 
of $\overline{\textrm{P}}$ANDA. The barrel EMC (Barrel) and both, forward (FwEndCap)
and backward (BwEndCap) EMC end-caps are shown.
{\bf{b)}} E/p versus track momentum for electrons and pions in the momentum range between 0.3~GeV/c 
and 5~GeV/c.}
\end{center}
\end{figure}

\vspace*{-0.3cm}
\hspace*{-0.4cm} To enable a rich physics program with the 
$\overline{\verb+P+}$\verb+ANDA+ detector,
all components have to be carefully designed and tuned. 
The 
$\overline{\verb+P+}$\verb+ANDA+ detector, with its components, is presented
in Fig.~\ref{fig1} and consists of two parts, a $\underline{\textrm T}$arget 
$\underline{\textrm S}$pectrometer
(\verb+TS+) and a $\underline{\textrm F}$orward $\underline{\textrm S}$pectrometer (\verb+FS+). 
Precise tracking and $\underline{\textrm P}$article 
$\underline{\textrm {ID}}$entification (\verb+PID+)
are crucial to measure momenta of particles with a good detection efficiency.
This is achieved by exploiting a 
$\underline{\textrm M}$icro-$\underline{\textrm V}$ertex-$\underline{\textrm D}$etector
(\verb+MVD+), $\underline{\textrm S}$traw $\underline{\textrm T}$ube 
$\underline{\textrm T}$racker (\verb+STT+) or, alternatively
$\underline{\textrm T}$ime $\underline{\textrm P}$rojection $\underline{\textrm C}$hamber 
(\verb+TPC+), six $\underline{\textrm D}$rift 
$\underline{\textrm C}$hambers (\verb+DC+), a muon and Cherenkov 
(\verb+DIRC+$=\underline{\textrm D}$etection of $\underline{\textrm I}$nternally 
$\underline{\textrm R}$eflected $\underline{\textrm C}$herenkov) detectors,
marked in Fig.~\ref{fig1}.
The high resolution $\underline{\textrm{E}}$lectro$\underline{\textrm{M}}$agnetic 
$\underline{\textrm{C}}$alorimeter (\verb+EMC+) in the \verb+TS+ and
the \verb+FS+
is particular important to detect photons
in a very broad energy range, starting from a few tens of MeV up to several GeV, 
as expected in the reactions aimed for 
$\overline{\verb+P+}$\verb+ANDA+.  
All sub-detectors together will cover nearly the full solid angle. 
A very high precision anti-proton beam, $\Delta p/p$$\sim$10$^{-4}-$10$^{-5}$, 
together with a high luminosity, $L\sim$10$^{31}-$10$^{32}$ cm$^{-2}$s$^{-1}$, 
allow to measure very narrow, 50$-$100~keV, mass states of charmonium
and reactions with cross sections down to in the order of pb. 
The main requirements addressed to the tracking devices are 
a good momentum resolution at the percent level and the ability to handle high count rates. 
The \verb+MVD+ is the detector closest to the interaction point. 
This tracking device for charged particles is essential for a very 
precise determination of secondary decay vertices of short-lived particles.
A~spatial resolution below 100~$\mu$m is required and simulations show that 
11~$\mu$m can be achieved~[1].
The \verb+MVD+ is surrounded by the \verb+STT+ or the \verb+TPC+, 
which will be used for a main track reconstruction. 
The position resolution in~$x$ and~$y$ coordinates 
of about 150~$\mu$m and a re\-solution of 3~mm in $z$ direction is expected. 
The \verb+TPC+ or the \verb+STT+ will be combined with the 
sub-system consisting of 
$\underline{\textrm G}$as $\underline{\textrm E}$lectron 
$\underline{\textrm M}$ultiplier (\verb+GEM+) detectors to perform tracking 
in the \verb+FS+ at angles below 22$^\circ$. The \verb+GEM+s have rate capabilities 
thousand times larger than 
 achievable with drift chambers, therefore they can handle a 
higher flux of particles. 
The generator for background reactions (\verb+DPM+ [6]) 
was used to generate 
events at the momentum of 3.6~GeV/c and the 
reconstructed paths from the \verb+MVD+ combined with the \verb+TPC+ tracks are
presented in Fig.~\ref{fig2}.a. 
This graph shows a very good matching of track
segments coming from these two detectors.
One of the systems used for \verb+PID+ 
in $\overline{\verb+P+}$\verb+ANDA+  
is the \verb+DIRC+ detector,
used for identifying particles with momenta above 1 GeV/c.
The velocity information determined from the Cherenkov angle 
$\Theta=arccos(1/n\beta)$ of a charged particle, propagating 
in the medium with the refractive index $n$,
is combined with the momentum information from the tracking detectors.
Fig.~\ref{fig2}.b presents 
\vspace*{-0.2cm}
the identification of 10$^{4}$ particles generated 
with the \verb+DPM+ and reconstructed by the \verb+MVD+, \verb+TPC+ and \verb+DIRC+ 
detectors. 
A very good separation of charged particles, $\pi^{+}$, $\pi^{-}$ and protons, 
has been shown.
The \verb+EMC+ detector, Fig. \ref{fig3}.a, 
\vspace*{-0.2cm}
\vspace*{-0.5cm}
is mainly used for photon detection, but it also measures
the deposited energy of electrons, positrons, muons and hadrons.
Muons and hadrons lose only a certain fraction of their kinetic energy
by ionisation processes, while $e^{-}$ and $e^{+}$%
deposit their complete energy in the electromagnetic 
shower. This information can be used for \verb+PID+ 
as illustrated in Fig.~\ref{fig3}.b. 
The $\overline{\verb+P+}$\verb+ANDA+  
physics program relies hea\-vi\-ly on the capability 
to measure photons with high energy, position and, partly, time resolutions 
over a wide dynamic range.
Thus, a highly granular \verb+EMC+ detector will be employed 
with fast PWO scintillators having a short radiation length, $X_{0}=$0.89~cm,
a high energy resolution for photons and electrons, 1.5\%/$\sqrt{E(GeV)}$+1.3\%  
and a very good time resolution, $<$1~ns for deposited energy above 80 MeV,
obtained from a measurement with a realistic prototype equippped with
a 1 cm$^2$ Avalanche PhotoDiode (APD) and Sampling ADC readout [7].
In the final configuration the PWO crystals will be equipped with two
APDs of 1 cm$^2$ size each, thus gaining an improvement factor up to
$\sqrt{2}$ for the energy and time resolution.
Such photon detectors are needed to cope with high count rates and
with the proposed compact design of the \verb+TS+.

\vspace*{-0.1cm}
\section{The charmonium $\textrm h_{c}$ (1$^{1}P_{1}$) reconstruction}
\begin{figure}[ht]
\begin{center}
\includegraphics[width=0.9\textwidth]{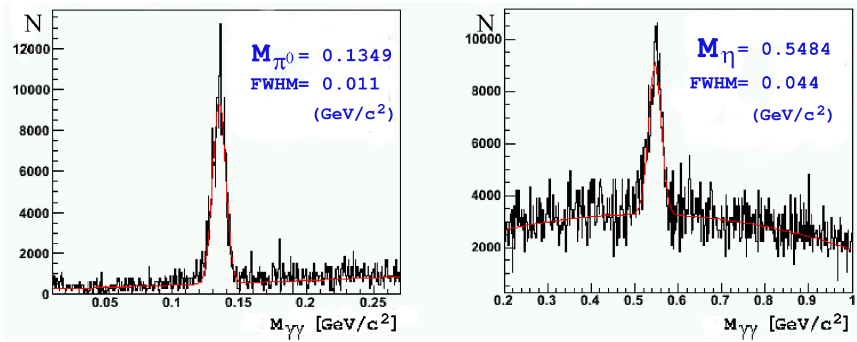}
\vspace*{-0.2cm}
\caption{The two-photon invariant mass spectra reconstructed in the EMC detector
from combinations of seven photons in the final state from chain (\ref{reaction}). 
A Gaussian fit combined with a second-order polynomial
is shown.}\label{fig4}
\end{center}
\end{figure}

\begin{figure}[ht]
\begin{center}
\begin{minipage}[t]{0.48\textwidth}
\includegraphics[width=5.5cm]{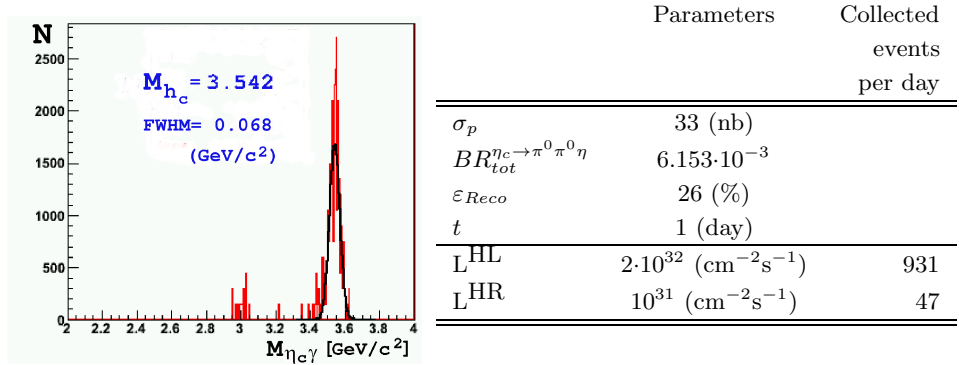}
\end{minipage} \hfill
\begin{minipage}[t]{0.48\textwidth}
\vspace*{-138pt}
\hspace*{-0.9cm}
\begin{tabular}{lcrrr}
& \fns{Parameters}  & \fns{Collected}  \\
&                   & \fns{events} \\
&                   & \fns{per day} \\
\hline
\hline
\fns{$\sigma_{p}$} & \fns{33~(nb)} &  \\
\fns{$BR_{tot}^{\eta_{c}\rightarrow\pi^{0}\pi^{0}\eta}$} & \fns{6.153$\cdot$10$^{-3}$} \\
\fns{$\varepsilon_{Reco}$} & \fns{26~(\%)}   &  \\
\fns{$t$}                                  & \fns{1~(day)}   &  \\
\hline
\fns{$\textrm L^{\textrm{HL}}$}   & \fns{2$\cdot$10$^{32}$~(cm$^{-2}$s$^{-1}$)} & \fns{931}  \\
\fns{$\textrm L^{\textrm{HR}}$}   & \fns{10$^{31}$~(cm$^{-2}$s$^{-1}$)}   &  \fns{47}  \\
\hline
\hline
\end{tabular}
\end{minipage}
\vspace*{-0.22cm}
\caption{\label{fig5}{\bf{Left:}} $\eta_{c}\gamma$ invariant mass reconstructed after 3$\sigma$ window 
cuts on $\pi^{0}$, $\eta$ and $\eta_{c}$ masses. 
{\bf{Right:}} Parameters used to calculate the collected number of events per day,
expected to reveal the signal of charmonium  $\textrm h_{c}$.
The calculated numbers of measured events/day are
presented in the lower part of the table.}
\end{center}
\end{figure}

\hspace*{-0.4cm} The invariant mass reconstruction requires a precise 
determination of the
energy deposited by photons and the opening angle between them, 
which can be achieved only by a highly granulated calorimeter;
the $\overline{\verb+P+}$\verb+ANDA+  
\verb+EMC+ fulfils this requirement.
The \verb+EMC+ plays a crucial role in the reconstruction of 
the singlet~{\small{(1$^{1}P_{1}$)}} state of charmonium,  $\textrm h_{c}$,
presented below. The following decay mode of the $\textrm h_{c}$:
\begin{equation}
\overline{\textrm{p}}+{\textrm{p}} \rightarrow {\textrm h_{c}} \rightarrow
\eta_{c} + \gamma \rightarrow (\pi^0+\pi^0+\eta) + \gamma \rightarrow 7\gamma
\label{reaction}
\end{equation}
has been studied in this work.
The chain~(\ref{reaction}) was generated with the \verb+EvtGen+ generator and
the invariant mass analysis was done using the \verb+Rho+ package~[1].
The two-photon invariant-mass spectra, reconstructed from combinations 
of seven photons in the final state are presented in Fig.~\ref{fig4}. Peaks which 
stem from $\pi^{0}$ and $\eta$ masses are found on top of  
\vspace*{-0.2cm}
\vspace*{-0.15cm}
a combinatorial background. A Gaussian function, 
describing the signal convoluted with a second-order polynomial,
describing the background, was fitted. Cuts
 with a window of 3$\sigma$ around the $\pi^{0}$ and $\eta$ 
mass peaks were applied in a further analysis and the background was reduced 
significantly.
With these cuts, a clean $\eta_{c}$ invariant mass spectrum was extracted
to finally identify the $\textrm h_{c}$ state, as demonstrated in Fig.~\ref{fig5}. 
The $\textrm h_{c}$ reconstruction efficiency, $\varepsilon_{Reco}$, 
obtained from the presented a\-na\-lysis was found to be 26\%.
The expected number of events to be measured during
the $\overline{\verb+P+}$\verb+ANDA+  
experiment was estimated
taking into account the production cross section, $\sigma_{p}$, 
which was calculated from the Breit-Wigner formula at the resonance 
ener\-gy for the process:
$\overline{\textrm{p}}{\textrm{p}}\rightarrow\textrm h_{c} \rightarrow
\eta_{c} + \gamma$~[1]. 
Also, the total bran\-ching ratio for the process $\eta_{c} \rightarrow \pi^{0} \pi^{0} \eta$, 
{\small{BR$_{tot}^{ \eta_{c} \rightarrow \pi^{0}\pi^{0}\eta}$}}, was used. 
In the table on the right hand
site of Fig.~\ref{fig5} the most important ingredients, which were used
in the analysis, are summarized.
The calculation was done for two modes available for $\overline{\verb+P+}$\verb+ANDA+:  
the \verb+high luminosity mode+, {\small{L$^{\textrm{HL}}$}},
with 
{\small{$\sigma_{p}/p\sim$10$^{-4}$}},
and the \verb+high resolution mode+, {\small{L$^{\textrm{HR}}$}}, 
with {\small{$\sigma_{p}/p\leq$4$\cdot$10$^{-5}$}} and the estimated numbers 
of collected events/day are presented in that table.
The obtained numbers are by factors of 5 and 10 larger for {\small{L$^{\textrm{HR}}$}} 
\vspace*{-0.55cm}
\vspace*{-1cm}
and {\small{L$^{\textrm{HL}}$}}, respectively,
than for other decay modes of the charmonium $\textrm h_{c}$ state,
such as 
$h_{c}\rightarrow\eta_{c}\gamma\rightarrow\Phi\Phi\gamma$ (2), described in~[1].
This is mainly due to the total bran\-ching ratio
{\small{BR$_{tot}^{\eta_{c}\rightarrow\pi^{0}\pi^{0}\eta}=6.153\cdot$10$^{-3}$}}, 
which is one order of magnitude larger than 
the total branching ratio of 
{\small{BR$_{tot}^{\eta_{c}\rightarrow\Phi\Phi}$=6.24$\cdot$10$^{-4}$}}.
The disadvantage might be an insufficient background suppression of the
decay channel~(\ref{reaction}) compared to the channel (2).
A complete background analysis of the competing channels as 
$\overline{\textrm p}\textrm p \rightarrow \pi^{0}\pi^{0}\pi^{0}$ 
and $\overline{\textrm p}\textrm p \rightarrow \pi^{0}\pi^{0}\eta$ is ongoing.

\vspace*{-0.1cm}
\section{Summary and conclusions}
With the $\overline{\verb+P+}$\verb+ANDA+  
detector a broad range of physics studies 
will be co\-vered. The precise measurement of the momentum and position of a particle
will be possible using highly tuned tracking, \verb+PID+ and \verb+EMC+ sub-detectors. 
The {\small{1$^{1}P_{1}$}} state of charmonium 
is one of several benchmark channels studied for 
$\overline{\verb+P+}$\verb+ANDA+,  
where masses and widths will be measured precisely. 
Recently, the $\textrm h_{c}$ was studied in $e^{+}e^{-}$ experiments
({\small{CLEO-c~[8]}}, {\small{BES3~[9]}}), 
but so-far only an upper limit for the total width, $\Gamma<$1~MeV,
has been published. In the $\overline{\textrm p}\textrm p$ annihilation
 experiment, such as $\overline{\verb+P+}$\verb+ANDA+,  
the exact value of $\Gamma$ 
will be measured.
In this work we analysed the decay of the 
$\textrm h_{c}\rightarrow\eta_{c}\gamma\rightarrow\pi^{0}\pi^{0}\eta$.
Preliminary studies show that the observation of this channel is feasible 
and further investigations of its background are under~way.

%
%
\vspace{0.3cm}
\tiny{[1] http://www-panda.gsi.de;}
\tiny{TPR (2005); PPR (2009), arXiv:0903.3905v1.}

\tiny{[2] S.~Spataro for the PANDA collaboration, Journal of Physics {\bf{119}}, 032035 (2008)}

\tiny{[3] http://root.cern.ch}

\tiny{[4] http://fairroot.gsi.de}

\tiny{[5] http://www.fluka.org}

\tiny{[6] A.~Capella et al., Phys.~Rept.{\bf{236}}, 225 (1994)}

\tiny{[7] M.~Kavatsyuk et al., IEEE Nuclear Science Symposium and Medical \\
\hspace*{1cm}Imaging Conference, 25-31.10.2009 Orlando, Florida}

\tiny{[8] S.~Dobbs et al., Phys.~Rev.~Lett.{\bf{101}}, 182003 (2008); arXiv:0805.4599v1.}

\tiny{[9] http://bes3.ihep.ac.cn/}
%
%
\end{document}